\documentstyle[11pt,psfig]{article}
\begin{document}

\begin{flushright} 
Utrecht
THU-96/01\\ 
hep-th/9601064\\ 
\end{flushright} 
\vskip 1cm 
\begin{center}
{\Large\bf On the Origin of the\\Outgoing Black Hole Modes}\\ 
\vskip 1cm 
{\large Ted Jacobson} 
\vskip .5cm 
{\it Institute for Theoretical Physics, University of Utrecht\\
P.O. Box 80.006, 3508 TA Utrecht, The Netherlands}\\ 
and\\ {\it Department of Physics, University of Maryland\\
College Park, MD 20742-4111, USA}\\
	     {\tt jacobson@umdhep.umd.edu}

\end{center} \vskip 1cm

\begin{abstract}
The question of how to account for the outgoing black hole modes
without drawing upon a transplanckian reservoir at the horizon is
addressed. It is argued that the outgoing modes must arise via
conversion from ingoing modes. It is further argued that the
back-reaction must be included to avoid the conclusion that particle
creation cannot occur in a strictly stationary background. The process
of ``mode conversion" is known in plasma physics by this name and in
condensed matter physics as ``Andreev reflection" or ``branch
conversion". It is illustrated here in a linear Lorentz non-invariant model
introduced by Unruh. The role of interactions and a physical short
distance cutoff is then examined in the sonic black hole formed with
Helium-II.

\end{abstract}

\section{The transplanckian reservoir}

A fundamental problem in black hole physics is to account for the
origin of the outgoing modes. In ordinary field theory these modes
arise from a reservoir of arbitrarily high frequency, short wavelength,
degrees of freedom (in a free-fall frame) propagating just outside the
event horizon and exponentially redshifting until they finally escape.
There are good reasons to doubt the existence of such a transplanckian
reservoir however.  Yet without this reservoir to draw upon, it seems
that a short time after the formation of a black hole there would be a
dearth of outgoing modes\cite{Jaco1}. Perhaps this is just what
happens.  But it would produce a catastrophic breakdown of the usual
vacuum structure outside a black hole horizon, would preclude the
existence of Hawking radiation, and would invalidate any semiclassical
analysis of black holes.  While this possibility can perhaps not (yet)
be ruled out conclusively by observations, it seems terribly unlikely.
Underlying this paper is the assumption that the outgoing modes do
indeed exist, and the problem is to account for them without drawing
upon a reservoir of transplanckian degrees of freedom.

It is easiest to describe the problem in free field terms, where a
``mode" of the field has an autonomous identity, but the problem also
exists for interacting fields. One way to describe it is to think of
the interacting theory perturbatively. Then the issue is that, in
ordinary field theory, the value of low energy observables far from the
hole at late times depends on transplanckian features of the propagator
near the horizon\cite{FredHaag}.  Nonperturbatively, one can view the
field theory as a set of coupled equations for the correlation
functions. In this framework, the operator equations of motion
presumably still imply that such observables depend on the
transplanckian structure of the correlation functions just outside the
horizon.

Two good reasons to doubt the existence of a transplanckian reservoir
are (i) field theory divergences and (ii) divergence of black hole
entropy in ordinary field theory.  It is generally believed that the
short distance divergences of quantum field theory are due to an
unphysical assumption about the short distance physics, namely that
there are  an infinite number of degrees of freedom localizable in any
volume, no matter how small that volume may be.  The existence of these
degrees of freedom also leads in general to a divergent contribution to
the entropy of a black hole. This can be viewed in terms of
entanglement entropy\cite{Sork,Bombetal,FrolNovi}, or thermal entropy
of acceleration radiation\cite{tHooft}, or simply in terms of the
renormalization of Newton's constant (and the coefficients of curvature
squared terms in the effective
action)\cite{SussUglu,Jacoindu}.\footnote{In fact it seems that due to
effects of ``curvature coupling", the entanglement and thermal
entropies are not identical. It is the ``thermal entropy" that appears
more fundamental when it is recognized as just one contribution to the
total entropy expression resulting from one piece of the low energy
effective gravity action (see, e.g., \cite{Frol-rev}).} In this last
guise it appears that the divergence of black hole entropy is merely
one aspect of the divergence problem of quantum field theory.  The cure
for this problem is to reject the assumption that there exists an
infinite density of localizable states.

\section{Nonlocality and Lorentz non-invariance}

Local Lorentz invariance requires that there be an infinite number of
degrees of freedom in any volume, no matter how small that volume may
be. To evade the conclusion that the density of states is infinite it
seems that one must give up either locality or Lorentz invariance.
Perturbative string theory is an example of a nonlocal theory in which
Lorentz invariance is maintained but the the density of localizable
states is finite (in fact zero)\cite{Suss-lor}.  As currently
formulated, string theory actually has many {\it more} states than
ordinary field theory (see, e.g. \cite{Pol-rev}), however each state is
fundamentally completely nonlocal. (There exist conjectures that this
is a condensed phase of a more fundamental theory that has a truly
finite density of states\cite{AticWitt}.) The role of this nonlocality
in accounting for the outgoing string states around a black hole has
not yet to my knowledge been clarified (see however
\cite{Suss-prl,Suss-lor,Engletal}).

The other possibility is to abandon Lorentz invariance rather than
locality.  It might be considered foolish to question Lorentz
invariance.  After all, we certainly have no observations that
challenge its validity, and indeed most physicists seem inclined to
believe that it is an {\it exact} symmetry of nature.  However, it
should be remembered that the Lorentz group is noncompact, so it would
take observations at arbitrarily large boosts to confirm Lorentz
invariance. (At present the upper bound to the boost factors that have
been probed is around $\gamma\sim 10^{12}$ in transforming to the
center of mass frame in cosmic ray collisions.) Of course if Lorentz
invariance is only an approximate symmetry it will be necessary to
explain why it appears to hold so well where it has been tested.  In
this connection it is interesting to note that effective field theories
governed by a Lorentzian metric are known to occur in many condensed
matter systems where there is actually a preferred frame. If the
spacetime metric is of this nature, and if the preferred frame is the
cosmic rest frame, then it is clearly relevant that the earth is
essentially at rest in this frame.  However, one would still have to
explain why we do not see several different effective Lorentz metrics
for different low energy fields. Of possible relevance is the
observation\cite{ChadNiel,NielNino} that there is a tendency for the
effective metrics for different interacting fields to flow to a common
metric in the infrared.

In this paper we will consider the consequences of Lorentz
non-invariance for the issue of the origin of the outgoing modes. These
considerations may also be relevant for a nonlocal theory such as
string theory, since non-locally realized Lorentz invariance might look
somewhat like plain old Lorentz non-invariance in some domain. We shall
see that there is a mechanism, operating within the Lorentz
non-invariant model theories considered, whereby the outgoing modes
originate from certain ingoing modes, by a counter-intuitive but rather
simple process.

\section{Conversion of ingoing to outgoing modes}

In this section we consider in general terms the question of how the
outgoing modes could possibly be restored in a theory with a finite
density of states.  This question was raised in Ref. \cite{Jaco1},
where it was proposed that a process of ``mode regeneration" must take
place, but the nature of this process remained obscure.  It was
conjectured that it necessarily involves {\it interactions}, and
perhaps corresponds to the time reverse of a decay process in which low
frequency outgoing modes fuse to form high frequency ones.  There are
(at least) two serious objections to this conjecture.  First, one still
needs to account for the low frequency outgoing modes, so not much has
been explained. The other objection, pointed out by L.
Ford\cite{Fordpc}, is that the outcome of such a process would
undoubtedly depend on the details of the interactions, for instance on
the coupling constants. The regeneration of the outgoing modes would
only be partial, which would be inconsistent with the standard vacuum
structure and would lead to violation of the generalized second law
because the hole could not emit the full thermal spectrum of Hawking
radiation. The only counter argument\cite{Jaco1} was to further
conjecture that the regeneration is like an equilibration process,
which yields the usual spectrum of vacuum fluctuations provided enough
time passes.  However this does not evade the first objection.

How many outgoing modes must be accounted for? If the black hole
background were static, the number would be infinite, since they could
come out at any time. A black hole left alone would presumably
evaporate in a time of order $M^3$ (in Planck units), however one can
in principle easily maintain the static background by sending energy
into the hole at a rate equal to the Hawking luminosity. Thus it seems
that one must be able to account for a truly infinite number of
independent outgoing modes.  Supposing that the ultimate density of
states is no greater than one per Planck volume, it follows that the
modes must originate in an infinite volume.\footnote{Allowing the black
hole to evaporate, the number is instead finite. Consider just the
$s$-waves, and divide the frequency range into intervals of a small
size $\epsilon$. Below the Planck frequency there are ${\epsilon}^{-1}$
frequency intervals, and each interval defines a wavepacket with time
spread $\sim {\epsilon}^{-1}$. In the evaporation time $M^3$ there are
therefore ${\epsilon}^{-1}(M^3/{\epsilon}^{-1})=M^3$ independent
$s$-wave modes that emerge. Since the high angular momentum modes have
little cross-section for absorption by the hole, the total number of
modes should also be of order $M^3$. Again assuming one Planck volume
per mode, this requires a total volume $M^3$.  That is, the minimum
distance from the horizon at which the modes can originate is of order
$M$.}

It seems that there are only two ways to account for these outgoing
modes without drawing upon a transplanckian reservoir:
\begin{enumerate} 
\item modes may emerge from the singularity and
propagate superluminally out across the horizon; 
\item ingoing modes
may be changed into outgoing ones.  
\end{enumerate} 
One can construct
dispersive linear field theory
models\cite{Unruh2,BMPS,Jaco-mex,CorlJaco} where either possibility
occurs. The first possibility would require the specification of
boundary conditions at the singularity, and does not seem as natural
as the second.  (It is not appropriate to reject  {\it a priori} the
first possibility, since we are questioning Lorentz invariance.) The
second possibility probably occurs for a sonic black hole\cite{Unruh1}
constructed with real atomic fluid, as will be discussed below.

Conversion from ingoing to outgoing modes is a phenomenon that is known
in other areas of physics. In plasmas it is called ``mode conversion",
which occurs when waves propagate in an inhomogenous
plasma\cite{Stix,Swanson}. In condensed matter it is called ``Andreev
relection", or ``branch conversion", which occurs in superfluid systems
where the order parameter is position-dependent, such as a
normal-superconductor interface\cite{Andr}, or a superflow or other
``texture" in superfluid $^3$He\cite{GreaLegg}.  In fact, a simple
analog of the outgoing black hole modes is the quasiparticle modes with
energy $\epsilon<\Delta$ propagating in a normal conductor away from an
interface with a superconductor with energy gap $\Delta$. The origin of
these modes is to be found in ingoing quasihole modes which undergo
Andreev reflection.

In the black hole case, certain ingoing short wavelength modes are
converted to outgoing short wavelength modes just outside the horizon,
which then redshift down to long wavelengths as they climb away from
the hole.  These ingoing modes are actually {\it outgoing} as viewed in
the free-fall frame of the black hole. Thus, from the viewpoint of a
free-fall observer, the conversion does not lead to a deficit of
ingoing modes.

Mode conversion from ingoing to outgoing modes may provide a
satisfactory mechanism for the mode regeneration. It can happen even in
{\it free} field theory, so it does not suffer form Ford's objection,
it presumably survives in the presence of interaction, and the
mechanism is sufficiently universal to account for the outgoing modes
in a wide range of theories.

\section{The stationarity puzzle}

If all outgoing wavepackets can be traced backwards in time to ingoing
ones that have never encountered the collapse process which formed the
black hole, then for them the black hole spacetime appears stationary.
Since Killing frequency is conserved, particle creation for them
appears impossible. How can particle creation occur in the presence of
a conserved Killing frequency?

The mere existence of a conserved Killing frequency is not the
problem.  For instance in deSitter space, a space of maximal symmetry,
ordinary quantum fields are excited even in the Killing
vacuum\cite{GibbHawk}. This can happen because the Killing frequency
does not define the relevant notion of particle states. A black hole
spacetime is asymptotically flat, however, and the Killing time agrees
at infinity with the relevant time for defining the asymptotic particle
states (assuming any preferred frame at infinity is the one in which
the black hole is at rest).  It would therefore appear there can be no
particle production by an eternal stationary black hole. The
annihilation operator $a_{\rm out}$ for an outgoing positive Killing
frequency wavepacket would be expressible in terms of the annihilation
operator $a_{\rm in}$ for an ingoing positive Killing frequency
wavepacket. With the standard in-vacuum boundary condition, $a_{\rm
in}|\Psi\rangle=0$, the number $N_{\rm out}|\Psi\rangle$ would
therefore vanish.

Three possible escape routes from this no-creation conclusion might be
imagined:  \begin{enumerate} \item the ingoing modes that give rise to
the outgoing modes do {\it not} originate at infinity; \item quantum
field theory breaks down at short distances; \item the back-reaction
destroys the Killing symmetry and decoheres positive Killing frequency
superpositions.  \end{enumerate} In the presence of a cutoff, as argued
above, the modes must originate in an infinite volume, so this escape
route (1) seems precluded due to an ``overcrowding" problem. The only
way out is if the modes were to emerge from the singularity, rather
than coming in from infinity.  Escape route (2) is that, as a
wavepacket is followed backwards in time out to infinity, it may
blueshift so far that the field theory description breaks down. After
that, the connection between ``Killing frequency at infinity" and norm
in Hilbert space might dissolve.  This escape route doesn't seem too
promising.  While it is plausible that the field theory description
breaks down, one would still expect to have a conserved Killing energy,
and this may be sufficient for a no-creation argument.

Escape route (3) is by far the most promising. In the real problem the
gravitational field is dynamical and couples to all other fields.  The
quantum evolution of the coupled system does not preserve the Killing
symmetry of the classical background. The high wavevector wavepackets
that form the incoming mode are presumably correlated to different
states of the gravitational field at short distances, and these states
lack the Killing symmetry of the background. Furthermore, there is the
effect of decoherence. Neglecting back-reaction, the out wavepacket
evolves backwards in time to a sum of in wavepackets $\phi_+ + \phi_-$,
where $\phi_+$ and $\phi_-$ have positive and negative norms
respectively.  It is only the {\it sum} that has positive Killing
frequency.  If the back-reaction is taken into account, the states
corresponding to these two wavepackets are presumably correlated to
different, perhaps orthogonal, states of the gravitational field. In
this case it would be incorrect simply to sum them together in forming
the annihilation operator $a(\phi_+ + \phi_-)=(\phi_+ +
\phi_-,\hat{\Phi})$ (where $(,)$ is the conserved norm and $\hat{\Phi}$
is the field operator), even if the Killing symmetry were preserved.

It seems quite plausible that some combination of these two effects of
the back-reaction can evade the stationarity paradox. Indeed there has
already been much interesting work on the consequences of the
back-reaction on the phase of the Bogoliubov coefficients in the
Hawking effect, and the existence of the kind of decoherence effect
being invoked here \cite{backreac}. Particularly encouraging is
Parentani's work which establishes that decoherence due to the
back-reaction gives rise to an energy flux from a ``uniformly"
accelerated particle detector\cite{Pare-det} or mirror\cite{Pare-mir},
even though such systems radiate no energy in the background field
approximation\cite{FullDavi,Grove}. (It should be admitted however that
in these systems there is a number expectation value for the outgoing
modes even in the background field approximation, whereas in our black
hole scenario even the number vanishes.) Our conclusion is that it
seems possible in principle to produce the Hawking radiation even if
outgoing modes arise from modes that are ingoing after the collapse
that formed the black hole.

\section{Dispersive models}

In this section we discuss the process of mode conversion in a free
field model that violates local Lorentz invariance on account of the
presence in the action of higher spatial derivative terms in the
free-fall frame of a black hole. For simplicity the model is restricted
to two spacetime dimensions. The model arose from considerations of
Unruh's sonic black hole analog\cite{Unruh1}, which we now briefly
describe.

In Unruh's analogy, the perturbations of a stationary background fluid
flow are quantized. If the background flow goes supersonic there is a
``sonic horizon", from beyond which sound cannot escape. Equating the
sound field to a massless free field, Unruh argued that the sonic
horizon will emit thermal Hawking phonons at a temperature $v'/2\pi$,
where $v'$ is the gradient of the background velocity field at the
horizon. One can begin to take into account the atomic nature of a real
fluid via the departure from linearity of the dispersion relation
$\omega(k)$ for phonons\cite{Jaco1}.  The key point is that the slope,
which gives the group velocity of wavepackets, is not constant, and in
fact initially decreases as $k$ increases. This dispersion relation
holds in the co-moving frame of the fluid, and leads to the phenomenon
of mode conversion from ingoing to outgoing modes as demonstrated by
Unruh\cite{Unruh2}. The model considered in \cite{Unruh2} is not a real
fluid but rather free field theory with higher spatial derivative terms
designed to produce a non-linear dispersion relation.

Unruh's model can be reinterpreted without the fluid flow
interpretation as a theory of a free quantum field in a black hole
spacetime. Let us describe the model in a slightly generalized
form\footnote{All of my understanding of this model has been developed
in collaboration wth S. Corley.}\cite{Jaco-mex,CorlJaco}.  The model
consists of a free, hermitian scalar field propagating in a two
dimensional black hole spacetime. The dispersion relation for the field
lacks Lorentz invariance, and is specified in the free fall frame of
the black hole, that is, the frame carried in from the rest frame at
infinity by freely falling trajectories.\footnote{A related model was
invented in Ref. \cite{BMPS}, which imposes the altered dispersion
relation in Eddington-Finkelstein coordinates rather than free-fall
coordinates.} Let $u^\alpha$ denote the unit vector field tangent to
the infalling worldlines, and let $s^\alpha$ denote the orthogonal,
outward pointing, unit vector, so that $g^{\alpha\beta}=u^\alpha
u^\beta-s^\alpha s^\beta$.  The action is assumed to have the form:
\begin{equation}
S={1\over2}\int d^2x \, \sqrt{-g} g^{\alpha\beta} {\cal D}_\alpha\phi^* {\cal D}_\beta\phi,
\label{action}
\end{equation}
where the modified differential operator ${\cal D}_\alpha$ is defined by
\begin{eqnarray}
u^\alpha {\cal D}_\alpha &=& u^\alpha \partial_\alpha\\
s^\alpha {\cal D}_\alpha &= &\hat{F}(s^\alpha \partial_\alpha).
\end{eqnarray}
The time derivatives in the local free fall frame are thus left
unchanged, but the orthogonal spatial derivatives are replaced by
$\hat{F}(s^\alpha \partial_\alpha)$.  The function $\hat{F}$ determines
the dispersion relation.  Invariance of the action (\ref{action}) under
constant phase transformations of $\phi$ guarantees that there is a
conserved current for solutions and a conserved ``inner product" for
pairs of solutions to the equations of motion.

The black hole line element we shall consider is static and has the
form 
\begin{equation} 
ds^2=dt^2-(dx-v(x)\, dt)^2 
\end{equation} where
$v(x)$ is negative and increasing to the right, going to a constant
$v_0<0$ at infinity. The black hole is at rest in these coordinates if
$v_0$ vanishes.  This is a generalization of the Lema\^{\i}tre line
element for the Schwarzschild spacetime, which is given by
$v(x)=-\sqrt{2M/x}$ (together with the usual angular part).
$\partial_t$ is a Killing vector, of squared norm $1-v^2$, and the
event horizon is located at $v=-1$.  The curves given by $dx/dt=v$ are
timelike free fall worldlines which are at rest (tangent to the Killing
vector) where $v=0$.  Since we assume $v<0$ these are {\it ingoing}
trajectories.  $v$ is their coordinate velocity, $t$ measures proper
time along them, and they are everywhere orthogonal to the constant $t$
surfaces.  We shall refer to the function $v(x)$ as the {\it free-fall
velocity}.  The asymptotically flat region corresponds to
$x\rightarrow\infty$.

In terms of the notation above, the orthonormal basis vectors adapted to the
free fall frame
are given by $u=\partial_t+v\partial_x$ and $s=\partial_x$, and
and in these coordinates $g=-1$. Thus the action (\ref{action}) becomes
\begin{equation}
S={1\over2}\int dtdx\,
\Bigl(|(\partial_t+v\partial_x)\phi|^2 -|\hat{F}(\partial_x)\phi|^2\Bigr).
\end{equation}
If we further specify that $\hat{F}(\partial_x)$ is an odd function of
$\partial_x$,
then integration by parts yields the field equation
\begin{equation}
(\partial_t+\partial_x
v)(\partial_t+v\partial_x)\phi=\hat{F}^2(\partial_x)\phi.
\label{eom}
\end{equation}

The behavior of wavepackets in this model can be understood
qualitatively as follows.  Assume a solution to the field equation
(\ref{eom}) of the form $\phi = e^{-i \omega t} f(x)$ and solve the
resulting ODE for $f(x)$ by the WKB approximation. That is, write $f(x)
= \exp(i\int k(x)\,dx)$ and assume the quantities $\partial_x v$ and
$\partial_x k/k$ are negligible compared to $k$.  The resulting
equation is the position-dependent dispersion relation
\begin{equation}
(\omega - v(x)k)^{2} = F^{2}(k),
\label{disp'}
\end{equation}
where $F(k)\equiv-i\hat{F}(ik)$.
This is just the dispersion relation in the local free-fall frame, since the 
free-fall frequency $\omega'$ is related to the Killing frequency $\omega$ by
\begin{equation}
\omega'=\omega-v(x)k.
\label{otrans}
\end{equation}

The choice of the function $F(k)$ completes the definition of the model. The
ordinary wave equation corresponds to $F(k)=k$.  Expanding in $k$, one has
\begin{equation}
F(k)=k-k^3/k_0^2+\dots
\end{equation}
(assuming reflection invariance). The wavevector $k_0$ characterizes
the scale of `new physics'. The only qualitative choice being made here
is that the cubic term is negative. In many condensed matter systems
the dispersion relation behaves in this way, and it is necessary in
order that (for an interacting field) the excitations be stable against
decay into longer wavelength ones\cite{Pita59}. The group velocity in
the free-fall frame is $dF/dk=1-3k^2/k_0^2+\dots$, which decreases
initially (at least) as the wavevector grows. This decrease in the
group velocity is the essential feature for us.

\begin{figure}[tb]
\centerline{
\psfig{figure=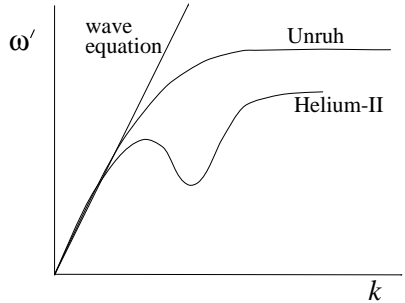,angle=-90,height=4cm}}
\caption{Sketch of dispersion relations $\omega'(k)$ for the wave equation,
the Unruh model, and Helium-II at zero temperature and pressure. 
$\omega'$ is the frequency in the free-fall frame (\ref{otrans}).}
\label{disps}
\end{figure}

Unruh's choice\cite{Unruh2} for the function $F(k)$ was
\begin{equation}
F_{\rm Unruh}(k)=k_0\{\tanh[(k/k_0)^n]\}^{1\over n}
\label{FUN}
\end{equation}
for various integers $n$. For every $n$ this has the feature that the
group velocity vanishes for large wavevectors and the frequency
approaches a maximum.  This dispersion relation is in some ways like
the dispersion relation for superfluid helium-4, with the roton minimum
taken out.  (Other choices for $F(k)$ are studied in Ref.
\cite{CorlJaco}, but these will not be discussed here.) This dispersion
relation is plotted in Fig. \ref{disps} along with the dispersion
relations for the ordinary wave equation and for quasiparticle
excitations of superfluid helium-4.

The dispersion relation is useful for understanding the motion of
wavepackets that are somewhat peaked in both position and wavevector.
The change in position can be found by integrating the group velocity
$d\omega/dk=v(x)+dF/dk$ while satisfying the dispersion relation
(\ref{disp'}).  A graphical method we have employed is described in
Refs.  \cite{Jaco-mex,CorlJaco}.  The same method was used by
BMPS\cite{BMPS}, who also found a Hamiltonian formulation for the
wavepacket propagation using Hamilton-Jacobi theory.

The model with the dispersion relation (\ref{FUN}) was solved in Ref.
\cite{Unruh2} by numerical integration of the PDE (\ref{eom}), and the
results can be reproduced qualitatively by the WKB methods. Propagating
a low wavevector outgoing wavepacket backwards in time, the horizon is
approached, the wavevector blueshifts to something of order $k_0$, and
the group velocity drops to zero in the static frame.  Mode conversion
occurs near the stopping point\cite{Jaco-mex,CorlJaco}, and the
wavepacket moves back away from the horizon with large positive and
negative wavevector components. (Forwards in time these components are
ingoing in the static frame but outgoing in the free-fall frame. This
happens because their group velocity in the free-fall frame is smaller
in magnitutde than $v(x)$ on account of the flattening of the
dispersion curve at large wavevectors.) These have positive and
negative free-fall frequencies and (therefore) positive and negative
norms respectively.  The magnitude of the component wavevectors grows
without bound as the wavepacket moves outward where $v(x)$ decreases,
and the asymptotic group velocity is just $v(x)$.  Were $v(x)$ to drop
to zero, the wavevector would diverge.  To avoid dealing with this one
can impose the in-vacuum in the free-fall frame at nonzero $v(x)$.
(This was done in \cite{Unruh2} and \cite{CorlJaco}.) No matter how
small $v$ gets, the same result is obtained for the (negative) norm of
the negative wavevector piece, which is the Bogoliubov coefficient that
determines the particle creation amplitude for the outgoing
wavepacket.  Thus, even though the difference between the free-fall and
Killing frames is going to zero as $v$ goes to zero, the wavevector is
diverging in such a way that the wavepacket always maintains a negative
free-fall frequency part of the same, negative, norm.

{}From this analysis we see that the Unruh model, while it entails a
strict cutoff in free-fall frequency, involves in an essential way
arbitrarily high wavevectors, i.e., arbitrarily short wavelengths.
Insofar as we wish to explore the consequences of a fundamental short
distance cutoff, this is an unsatisfactory feature of the model. The
outgoing modes emerging from the black hole region still arise from
arbitrarily short wavelength modes, albeit ingoing ones.

\section{Helium-II sonic black hole} 
For a physical model with a strict
cutoff let us consider the behavior of liquid helium at zero
temperature. This is of course an interacting system, so is not nearly
as simple as the Unruh model. Nevertheless, is is possible to make some
reasonable conjectures based on the form of the quasiparticle spectrum
in Fig. \ref{disps}.  (To avoid the need to consider interactions, one
might study instead field theory on a lattice falling into a black
hole.) In discussing the helium model, it is natural to go back to
Unruh's original sonic analogy and think of the free-fall velocity
$v(x)$  as the velocity of the background fluid flow. To begin with,
let us ignore the existence of interactions and just follow modes as if
they were free.

\begin{figure}[tb]
\centerline{
\psfig{figure=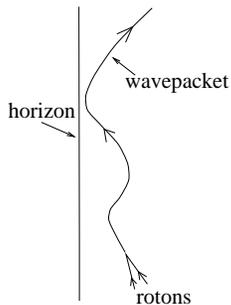,angle=-90,height=4cm}}
\caption{Sketch of the history of an outgoing wavepacket
propagated with the dispersion relation for Helium-II in (Fig. \ref{disps}).}
\label{dance}
\end{figure}

In \cite{Jaco1} it was argued that in the helium model a long
wavelength outgoing wavepacket, traced backward in time, would come to
rest at an ``effective horizon" where the co-moving group velocity and
fluid velocity are equal and opposite. For this part of the process,
the difference between the helium dispersion relation and that of the
Unruh model (\ref{FUN}) is irrelevant, so Unruh's results\cite{Unruh2},
as well as those of \cite{BMPS} and \cite{CorlJaco}, show that this
expectation is incorrect. Rather, the blueshifting continues, the group
velocity continues to drop, and the wavepacket is swept back out away
from the sonic horizon as a superposition of positive and negative
wavevector packets. Now let us continue to follow the progress of, say,
the positive wavevector part, backward in time, using the dispersion
relation of liquid helium.  The idea is sketched in Fig. \ref{dance}.
(The behavior of the negative wavevector part is similar.) The packet
will go over the first maximum of the dispersion curve, at which point
its co-moving group velocity changes sign, which only pushes it away
from the horizon even faster.  Eventually, however, it approaches
another turn around point, near the roton minimum, where the free-fall
frequency line becomes tangent once again to the dispersion curve. It
seems reasonable to suppose that what happens here is another reversal
of direction, with the wavepacket continuing along the dispersion curve
and falling back towards the horizon. As the wavevector rises and group
velocity falls one more tangency point will be reached, where there is
presumably one final reversal of direction. After that (still backward
in time), the wavepacket heads back away from the horizon with still
blueshifting wavevector and finally runs off the end of the
quasiparticle spectrum.  The quasiparticle spectrum terminates when a
decay channel into two rotons opens up, at twice the momentum and
energy of the roton minimum\cite{Pita59,Khal}.  Therefore, it appears
that our outgoing mode originates as some two roton mode (with
vanishing co-moving group velocity) that is swept in by the flow. That
is, the number operator for outgoing long wavelength phonons is
dynamically related to the 4-point function for ingoing rotons.

We were forced to incorporate the interactions when the end of the
quasiparticle spectrum was reached, but of course the interactions play
some role all along that we have ignored. One way to think about this
is to ask about the stability of quasiparticles. If other decay
channels are kinematically available, then presumably these are mixed
in to the evolution of the vacuum correlation functions.  There are
indeed two regions of the quasiparticle spectrum that are unstable
besides the endpoint.  First, there is  ``anomalous dispersion" at low
wavevectors, where $d^2\omega/dk^2$ is actually positive rather than
negative\cite{Grif}, which leads to a finite phonon
lifetime\cite{Pita59,Khal}.  Second,  past the roton minimum, there is
a region where the group velocity just reaches the velocity of
sound\cite{WoodCowl}, leading to phonon emission. The existence of
these processes presumably implies a mixing of the multipoint vacuum
correlation functions.
Thus an outgoing phonon mode really arises from a superposition of
various numbers of phonon modes, each of which ultimately arises from
ingoing multi-roton modes.

In summary, we thus conjecture the following forward in time behavior.
As particular superpositions of multi-roton modes of the superfluid are
swept in towards the horizon, the interactions and velocity gradient
conspire to turn them (after some dancing around near the horizon) into
outgoing phonon modes which are in (or very nearly in) their ground
state. At this point the Hawking effect is responsible for populating
them thermally as they climb away from the horizon.  The negative
energy flux across the horizon which is required by energy conservation
must leave the superfluid in a state with lower energy density than the
homogeneous superfluid ground state.

In fact, the true behavior of helium in the presence of a sonic horizon
is complicated by the instability towards vortex and roton creation. In
this connection it is interesting to note that a periodic roton
condensate will develop\cite{Pita84} at a critical velocity $v\sim 60$
m/s which is much less than the velocity of sound (238 m/s) in helium.
Perhaps this phase transition is essential to understanding the energy
balance in the presence of the sonic Hawking effect in helium.

\section{Information loss}

Finally, we point out that in the models considered here, it appears
that the presence of a cutoff and violation of Lorentz invariance does
not change the picture with regard to information loss in black hole
evaporation. The created particles still have a ``partner"\cite{BMPS},
which falls down into the singularity, to whom they are
correlated\footnote{It should be noted that the usual partner,
obtained\cite{Unruh-notes} by reflecting an out mode across the
horizon, is another mode with transplanckian pedigree in ordinary field
theory.  In the dispersive models, this partner apparently\cite{BMPS}
also arises from an ingoing (negative Killing frequency) mode.}.  We
see no mechanism for recovering the information in those correlations
based on these models.

\section{Acknowledgments} 
I am grateful to S. Corley for countless
discussions that have helped me in my effort to understand the problems
discussed here. I would also like to thank R. Brout, S. Massar, M.
Ortiz, R. Parentani, F. Skiff, Ph. Spindel, G. `t Hooft, W.G. Unruh,
and G. Volovik
for helpful discussions. This work was supported in part by NSF Grant
PHY94-13253 and the University of Utrecht.

\end{document}